\documentclass[journal,twoside,final]{IEEEtran}
\usepackage{cite}
\usepackage{amsmath}
\usepackage{amssymb}
\usepackage{amsthm}
\usepackage{thmtools}
\usepackage{graphicx}

\newtheorem{theorem}{Theorem}
\newtheorem{prop}{Proposition}
\theoremstyle{definition}
\newtheorem{definition}{Definition}
\declaretheorem[style=remark,qed=$\blacktriangleleft$]{example}

\title{Zipper Codes}
\author{Alvin Y. Sukmadji, \emph{Graduate Student Member, IEEE},
Umberto Mart\'{i}nez-Pe\~{n}as, \emph{Member, IEEE}, and\\
Frank R. Kschischang, \emph{Fellow, IEEE}\thanks{Submitted for
publication on March 16, 2022; revised on June 15, 2022; accepted on July 19, 2022.\newline
A.~Y.~Sukmadji and F.~R.~Kschischang are with the Edward S.\ Rogers Sr.\ Department
of Electrical \& Computer Engineering, University of Toronto, Canada.
U.~Mart\'{i}nez-Pe\~{n}as is with the Institute
of Mathematics, University of Valladolid, Spain.  Emails:
alvin.sukmadji@mail.utoronto.ca, umberto.martinez@uva.es,
frank@ece.utoronto.ca.\newline This paper was presented in part at
the 16th Canadian Workshop on Information Theory, June 2019 \cite{sukmadji}.}
}
\begin{document}

\maketitle
\begin{abstract}
Zipper codes are a framework for describing spatially-coupled
product-like codes. Many well-known codes, such as staircase codes and
braided block codes, are subsumed into this framework. New types of
codes such as tiled diagonal and delayed diagonal zipper codes are
introduced along with their software simulation results. Stall patterns
that can arise in iterative decoding are analyzed, giving a means of
error floor estimation.
\end{abstract}


\section{Introduction}
\label{sec:intro}
\IEEEPARstart{Z}{ipper} codes \cite{sukmadji,sukmadjithesis} represent a
framework for describing spatially-coupled product-like error-correcting
codes that are widely implemented in optical communications systems.
Zipper codes encompass many well-known spatially-coupled codes such as
staircase codes \cite{smith}, braided block codes \cite{feltstrom-bbc},
diamond codes \cite{baggen}, continuously-interleaved
Bose-Chaudhuri-Hocquenghem (BCH) codes \cite{coe}, swizzle codes
\cite{northcott}, oFEC \cite{humblet,g7093}, and spatially-coupled turbo
product codes \cite{montorsi}.

The general structure of zipper codes is closely related to that of
spatially-coupled generalized low-density parity-check (SC-GLDPC) codes
\cite{costello-scgldpc,mitchell}, the main difference being that the
latter are designed to operate under soft-decision decoding, whereas the
former are designed to be decoded using low-complexity, iterative,
algebraic, hard-decision decoding algorithms.
In applications such as optical transport networks, where
data throughputs are beginning to approach 1~Tb/s per channel,
it is vital for the decoder hardware to be
very energy-efficient, as every pJ of energy spent per decoded
bit translates, at a throughput of 1~Tb/s, to a power consumption of 1~W.
As a result, spatially-coupled codes with hard-decision
decoding have gained in popularity due to their lower energy consumption
(per decoded bit)
when compared to codes with soft-decision decoders
\cite{jian,lentmaier15,weiner,ou,lee13}.

In some applications, zipper codes are of interest as outer codes in
concatenated coding schemes \cite{barakatain20,stern,barakatain21,tayyab},
particularly for high code rates, but they have also been considered as
inner codes \cite{tian,800GMSA}.  Efficient hardware implementation of
zipper codes is considered in \cite{xie}, where a decoder achieving a
throughput of $962$ Gbps operating at $500$ MHz clock frequency is
reported. Other related hardware implementations include
\cite{fougstedt,truhachev21}.

This paper provides a general overview of zipper codes, organized as
follows.  The structure and main ingredients of a zipper code are
discussed in Sec.~\ref{sec:structure}. Several important code families
that can be described as zipper codes are given in
Sec.~\ref{sec:examples}.  Software simulation results for a few example
zipper code designs are given in Sec.~\ref{sec:designexamples}. Finally,
stall patterns of zipper codes are
discussed and analyzed in Sec.~\ref{sec:stallpatternanalysis}.

Throughout this paper, we will assume that all codes are defined over
the binary field $\mathbb{F}_2=\{0,1\}$; however, analogous formulations
can be made for the non-binary case.  For any positive integer $q$, we let
$[q]=\{0,1,\ldots,q-1\}$.  The cardinality (number of elements) of a
finite set $A$ is denoted as $|A|$.  The natural numbers are given as
$\mathbb{N} = \{ 0, 1, 2, \ldots \}$.

\section{Structure of Zipper Codes}
\label{sec:structure}

\subsection{Constituent Codes, Buffer, Zipping Pair}

Let $\mathcal{C}_{0},\mathcal{C}_{1},\ldots$ be any
sequence of binary linear \emph{constituent codes} indexed by $i \in
\mathbb{N}$, where the $i$th code has length $n_i$ and dimension $k_i$.
We assume, without loss of generality, that each codeword
of $\mathcal{C}_i$ is composed of $k_i$
\emph{information symbols} located in a fixed
set of positions (an \emph{information set}) indexed by $I_i \subseteq [n_i]$,
and $n_i-k_i$ \emph{parity symbols} located
in the (complementary) positions indexed by $[n_i] \setminus I_i$.
In practice, we will usually order the positions of $\mathcal{C}_i$
so that $I_i = [k_i]$, in which case $\mathcal{C}_i$
admits a \emph{systematic encoder} which places $k_i$ information
symbols directly into the first $k_i$ codeword positions and which places
$n_i-k_i$ parity symbols into the last $n_i-k_i$ codeword positions.

A \emph{buffer} associated with a given sequence of constituent codes is
any sequence of binary row vectors $\boldsymbol{c} = \boldsymbol{c}_0,
\boldsymbol{c}_1, \ldots$ such that $\boldsymbol{c}_i
\in \mathbb{F}_2^{n_i}$ for each $i \in \mathbb{N}$.  Thus, a buffer is
a sequence of rows, with the $i$th row having $n_i$ positions, possibly
(but not necessarily) forming a codeword of $\mathcal{C}_i$.  For any $i
\in \mathbb{N}$ and any $j \in [n_i]$ the $j$th symbol of the $i$th row
of $\boldsymbol{c}$ is denoted as $c_{i,j}$;  this symbol is said to have
\emph{position} $(i,j)$ within buffer $\boldsymbol{c}$.  For purposes of
initialization (i.e., to establish suitable boundary conditions), we
extend the set of buffer positions to allow for negative row indices by
defining $c_{i,j} = 0$ for all $i < 0$ and all $j \in \mathbb{N}$.  Thus
if we refer to a buffer symbol with a negative row index, that symbol
necessarily has value zero.  When $i < j$ we will say that row $i$ is
\emph{older} than row $j$ (or, equivalently, that row $j$ is
\emph{newer} than row $i$).

%%%%%%%%%%%%%% FIGURE 1 %%%%%%%%%%%%%%%%%%%
\begin{figure}[t]
\centering
\includegraphics{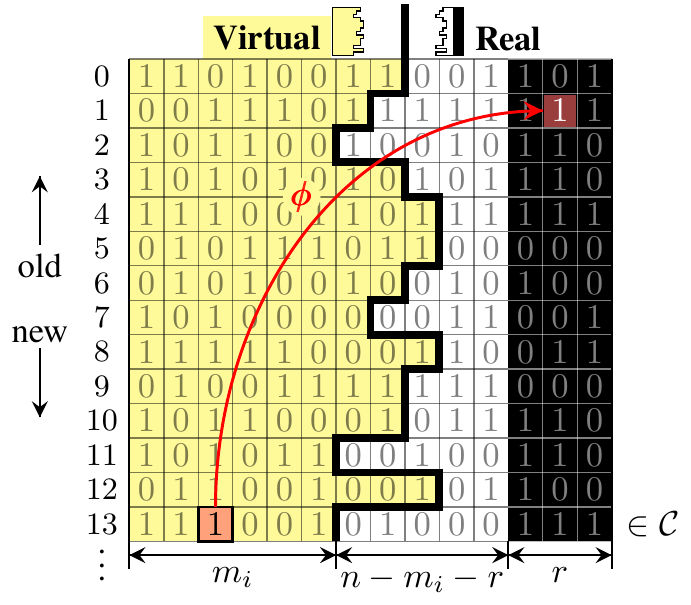}
\caption{Example of a zipper code with a systematically encoded $\mathcal{C}_i=\mathcal{C}$
constituent code with $n=n_i=14$ and $r=r_i=3$. Tiles in the shaded
region represent virtual symbols, while tiles in the unshaded and filled
regions represent real symbols. The filled regions show the location of
parity symbols.  The two tiles connected by arrows represent the two
coordinates prescribed by the interleaver map. In this example, we have
$\phi(13,2)=(1,12)$ and $c_{13,2}=c_{1,12}=1$. Each row is a codeword of
$\mathcal{C}$. Rows with lower row indices correspond to ``older'' rows
while those with higher indices are regarded as ``newer.'' The icons next to
``Virtual'' and ``Real'' at the top of the figure correspond to the shape
of the virtual and real buffers in the first 14 rows. The virtual and
real buffers are demarcated by a bold line.}
\label{fig:zipper}
\end{figure}
%%%%%%%%%%%%%%%%%%%%%%%%%%%%%%%%%%%%%%%%%%%%%

As illustrated in Fig.~\ref{fig:zipper},
we will \emph{partition} each row of a buffer $\boldsymbol{c}$ as follows.  For each
$i \in \mathbb{N}$, let $A_i \subseteq I_i$ be any (fixed) subset of the
information positions of $\mathcal{C}_i$, and let $B_i = [n_i] \setminus
A_i$ be the complementary set of positions.  The buffer positions
indexed by $A_i$ form the \emph{virtual positions} of the $i$th row (the
corresponding symbols are called virtual symbols), while those indexed
by $B_i$ form the \emph{real positions} (and the corresponding symbols
are called real symbols).  As we will soon see, only real symbols are
transmitted over the channel;  each virtual symbol of a constituent
codeword is a copy of a real symbol from some other constituent
codeword.  When $\mathcal{C}_i$ admits a systematic encoder, we will usually
take $A_i = [m_i]$ for some $m_i \leq k_i$, thus designating the first
$m_i$ positions as virtual positions.  Let
\[
A = \bigcup_{i \in \mathbb{N}} \{ (i,j): j \in A_i \} \text{ and }
B = \bigcup_{i \in \mathbb{N}} \{ (i,j): j \in B_i \}
\]
denote the positions of the virtual symbols and the real symbols,
respectively.  We refer to $(A,B)$ as a \emph{zipping pair}.  The set
$\{ c_{i,j} : (i,j) \in A \}$ is called the  \emph{virtual buffer}, and
the set $\{ c_{i,j}: (i,j) \in B \}$ is called the \emph{real buffer}.
The set $B^* = B \cup \{ (i,j): i \in \{ -1, -2, \ldots \}, j \in
\mathbb{N} \}$ is an index set for an \emph{extended real buffer} where
negative row indices are permitted.  As already noted, symbols located
in rows with a negative row index have value zero.

\subsection{Interleaver Map}

An \emph{interleaver map} associated with a zipping pair $(A,B)$ is any
function $\phi:A\rightarrow B^*$.  For each virtual position $(i,j) \in
A$, the interleaver map gives a real position $\phi(i,j) \in B^*$ from
which to copy a symbol.

We will often be interested in the row index from which a real symbol is
copied.  To this end, we define coordinate functions $\phi_1$ and
$\phi_2$ such that $\phi(i,j) = (\phi_1(i,j),\phi_2(i,j))$.  Thus
$\phi_1(i,j)$ returns the row index from which the virtual symbol at
position $(i,j)$ is copied.  If $\phi_1(i,j) < 0$, then the copied
symbol is necessarily zero.

For each real position $(i,j) \in B^*$, let $\phi^{-1}(i,j) = \{ (i',j')
\in A \colon \phi(i',j') = (i,j) \}$ denote the inverse image of $(i,j)$
under mapping by $\phi$.  Then $|\phi^{-1}(i,j)|$ gives the number of
virtual copies of the real symbol in position $(i,j)$.

Equipped with a sequence $\mathcal{C}_0, \mathcal{C}_1, \ldots$ of
constituent codes, a zipping pair $(A,B)$, and an interleaver map
$\phi$, we say that a buffer $\boldsymbol{c} = \boldsymbol{c}_0, \boldsymbol{c}_1, \ldots$ forms a valid
\emph{zipper codeword} if
\begin{enumerate}
\item $c_{i,j}=c_{\phi(i,j)}$ for all $(i,j) \in A$, and
\item $\boldsymbol{c}_i \in \mathcal{C}_i$ for all $i \in \mathbb{N}$.
\end{enumerate}
In other words, each virtual symbol of a valid zipper codeword must be a
copy of a real symbol as prescribed by the interleaver map $\phi$, and
each row of a valid zipper codeword must be a codeword of the
corresponding constituent code.

Fig.~\ref{fig:zipper} illustrates a zipper codeword in a zipper code
specified in terms of a sequence of constituent codes having fixed length
$n = n_i = 14$ and fixed dimension $k = k_i = 11$.  The constituent
codes are assumed to admit systematic encoders, so that parity symbols fall
into the last $r = n-k = 3$ positions in each row.  The first $m_i$
positions in each row are the virtual positions, while the last $n -m_i
$ positions are the real positions. Note that $m_i$ is permitted to vary
from row to row.  As illustrated, the interleaver map $\phi$ determines
the position at which to look up the value of a virtual symbol.

It is clear that the interleaver map determines the degree of coupling
among the various constituent codes.  We will usually focus on
interleaver maps that are \emph{periodic} and \emph{causal}.
\begin{definition}
An interleaver map $\phi$ is said to be \emph{periodic} with period $\nu
> 0$, or simply $\nu$-periodic, if $\phi(\nu + i, j) = (\nu,0) +
\phi(i,j)$ for all $(i,j) \in A$.
\end{definition}
To support a $\nu$-periodic interleaver map, we will require that the
zipping pair $(A,B)$ is also $\nu$-periodic in the sense that $n_{\nu +
i}=n_i$, and $A_{\nu+i} = A_{i}$ for all $i \in \mathbb{N}$.  Most often
we will then also have $\mathcal{C}_{\nu+i} = \mathcal{C}_i$ for all $i
\in \mathbb{N}$.
\begin{definition}
An interleaver map $\phi$ is said to be \emph{causal} if $\phi_1(i,j)
\leq i$ for all $(i,j) \in A$ and \emph{strictly causal} if
the inequality is strict.
\end{definition}
In other words, under a causal interleaver map, virtual symbols are
never copied from later rows.

Some important classes of zipper codes (such as staircase codes) have
the property that each real symbol is copied exactly \emph{once} into
some virtual position, i.e., the inverse image $\phi^{-1}(i,j)$ of every
real position $(i,j) \in B$ under mapping by $\phi$ is a singleton set
$\{ (i',j') \}$ for some $(i',j') \in A$.  In this situation we refer to
the interleaver map as being \emph{bijective}.

\subsection{Encoding Zipper Codes}

Recall that $I_i \subseteq [n_i]$ denotes an information set with $|I_i|
= k_i$ for the constituent code $\mathcal{C}_i$ of length $n_i$ and dimension $k_i$, and that $A_i
\subseteq I_i$ is an index set for the virtual positions of the $i$th
buffer row $\boldsymbol{c}_i$.  To encode $\boldsymbol{c}_i$, we assume that all rows $\boldsymbol{c}_j$ with
$j<i$ have already been encoded.  This will be true even for $c_0$,
since previous rows are then all-zero by assumption.  Under the
assumption of a causal interleaver map, the encoding of zipper codes is
easily accomplished by the following procedure:
\begin{enumerate}
\item Fill in the positions indexed by $B_i \cap I_i$ with message symbols.
\item Fill in the virtual symbols by duplicating symbols from the positions
prescribed by the interleaver map, i.e., for each $j \in A_i$, let $c_{i,j} = c_{\phi(i,j)}$.
(Since $\phi$ is causal by assumption, $c_{\phi(i,j)}$ is a symbol from
a row already encoded, or a symbol from the current row which was
input in the first step.)
\item Complete the row by filling in the parity symbols in positions indexed by
$[n_i] \setminus I_i$ using an appropriate encoder for $\mathcal{C}_i$, thereby
fulfilling the condition that $\boldsymbol{c}_i \in \mathcal{C}_i$.
\end{enumerate}
In the zipper code of Fig.~\ref{fig:zipper}, for example, the shaded
tiles in row $i$ are filled with symbols drawn from previous rows, the
unshaded tiles correspond to message symbols, and the filled tiles
correspond to parity symbols computed using the encoder for
$\mathcal{C}_i$.

\subsection{Code Rate}

We will always assume that only the symbols in the real buffer, i.e., in
positions indexed by $B$, are transmitted over the channel, in some
well-defined order from oldest to newest.  The symbols in the virtual
buffer, which are needed for purposes of encoding and decoding, are not
transmitted.  As usual, let $m_i = |A_i|$ denote the number of virtual
symbols in row $i$.  Under the assumption of that only the real buffer
is sent, the \emph{rate} of a zipper code is given as
\begin{equation}
R=\lim_{L \to \infty} \frac{\sum_{i=0}^{L-1} (k_i-m_i)}{\sum_{i=0}^{L-1} (n_i-m_i)}= \frac{\overline{k}-\overline{m}}{\overline{n}-\overline{m}},
\label{eq:rate}
\end{equation}
where, assuming the limits exist,
\[
\overline{k} = \lim_{L\to \infty} \sum_{i=0}^{L-1} \frac{k_i}{L},~
\overline{n} = \lim_{L\to \infty} \sum_{i=0}^{L-1} \frac{n_i}{L},~
\overline{m} = \lim_{L\to \infty} \sum_{i=0}^{L-1} \frac{m_i}{L}.
\]
For example, staircase codes with staircase block size $m\times m$ uses a fixed constituent code
of length $2m$ and dimension $k$. Thus, $\overline{n} = 2m$, $\overline{k} = k$,  and
$\overline{m} = m$, thereby achieving a rate $R = \frac{k}{m} - 1$
(assuming $\frac{k}{m} > 1$).

\subsection{Decoding}

Suppose that we transmit real symbols in positions indexed by $B$.  At
the receiver we can form a buffer $\boldsymbol{c}'$ by filling in the real positions
from symbols received at the output of the channel, and filling in
virtual positions by copying real received symbols as prescribed by the
interleaver map $\phi$.  We generally assume a hard-decision channel,
i.e., we assume that the channel outputs are elements of $\mathbb{F}_2$,
though it is possible to consider more general situations with various
amounts of reliability information in the form of erasures or
log-likelihood ratios.  The received buffer $\boldsymbol{c}'$ is not necessarily a
zipper codeword, since, due to channel noise or other impairments
causing detection errors, the rows aren't necessarily codewords of the
corresponding constituent codes.  Ideally, the aim of the decoder would
be to recover a valid zipper codeword while making as few changes to
$\boldsymbol{c}'$ as possible.  However, such optimal minimum Hamming distance
decoding is often too complicated to implement, so in practice some form
of \emph{iterative decoding} is used, making changes to the rows one at
a time using the constraints imposed only by the constituent code that
constrains that row, and usually revisiting each row a number of times.

Decoding is typically performed within a sliding window of $M$
consecutive rows $\boldsymbol{c}'_{i-M+1},\boldsymbol{c}'_{i-M+2},\ldots, \boldsymbol{c}'_i$ from the received
buffer.  Many of the virtual symbols in these rows will, however, be
copies of symbols from outside the sliding window (which, ideally, will
have been corrected by previous decoding iterations).  The decoder
operates by decoding the rows within the decoding window.  When the
decoder performs correction operations (flipping the value of one or
more bits in a row), all copies of the affected bits within the decoding
window (as determined by the interleaver map) will also need to be
flipped.  After a number of iterations have been performed, the sliding
window can be advanced (by one or, more usually, several rows), and the
corrected information symbols leaving the window can be delivered as the
decoder output.  Numerous variations of this basic scheme are possible. 

In one round of so-called \emph{exhaustive decoding}, every row in the
decoding window is visited (exactly once) by the corresponding
constituent decoder.  Each bit is decoded (in a round) as many times as
it appears in the window.  Exhaustive decoding can be performed in
serial fashion (one row at a time), or in parallel (usually under the
constraint that decoders operating in parallel don't operate on the same
bits).

In one round of so-called \emph{pipelined decoding}, only a subset of
the rows in the decoding window (every $L$th row, say, for some
parameter $L$) are visited by constituent decoders.  This method may be
faster than exhaustive decoding, but, depending on the interleaver map,
some bits may not be visited by a decoder at all.  Thus, pipelined
decoding is suitable only for certain types of interleaver maps that
ensure that as many bits as possible are visited by the constituent
decoders. Similar to exhaustive decoding, pipelined decoding can be done
in series or parallel.

Under both of these decoding methodologies, decoding can be performed
in multiple rounds,  until no more errors can be corrected or until some
maximum number of allowed decoding rounds is reached.

Additional strategies can also be applied to these decoding procedures.
In \emph{chunk decoding}, rather than advancing the sliding window one
row at a time, the window is advanced only after a specified number of
new rows, called a \emph{chunk}, is received.

In decoding with \emph{fresh/stale flags}, a ``fresh/stale'' status
indicator is maintained for each buffer row.  The indicator is set to
``fresh'' whenever a change is made to the row (for example, if the row
is newly arrived, or a bit has been flipped in that row due to a
correction elsewhere); otherwise, after a decoding it is set to
``stale''.  In each iteration, decoders can skip over stale rows (since
nothing has changed in that row since the previous iteration, and
therefore the row has either been corrected or has been determined to be
uncorrectable).

In decoding with \emph{periodic truncation}, a fraction of the message
positions are reserved to be set to known values (for example, zero).
The known symbols do not need to be transmitted, as they can be filled in
perfectly at the receiver.  One approach would be to alternate between
sending data in, say, $J$ consecutive rows, and not sending data (or,
more precisely, sending only parity symbols) in $\tau$ consecutive rows.
We call $J$ and $\tau$ the
\emph{transmission length} and \emph{truncation length}, respectively.
This method is a generalization of staircase code ``termination''
discussed in \cite{qiu}, and is a form of code ``shortening'' as defined
in coding theory.  The shortened code will have a lower code rate than the
zipper code from which it is derived.

In \emph{dynamic decoding}, the decoder is implemented with a number of
constituent decoders that can operate in parallel and that can be
assigned dynamically (according to a scheduling module) to the rows to
be decoded.  Equipping a decoding engine with the flexibility to assign
decoders to the rows where work needs to be done results in increased
hardware utilization (reduced computational idling).  Dynamic decoding
can achieve similar decoding performance with fewer computational
resources than without dynamic decoding \cite{huang}.

Although they fall outside the scope of this paper, a variety of
\emph{soft-decision} or \emph{soft-aided} decoder architectures are also
possible.  Among these, we mention anchor decoding \cite{hager2018},
``trusted symbol'' decoding \cite{senger}, soft-aided decoding
\cite{lei19,lei2021,sheikh21,sheikh21-ee}, and error-and-erasure
decoding \cite{sukmadjithesis,rapp}.

%%%%%%% FIGURE 2 %%%%%%%%%%%%%%%%%%%%%%%%%%%%%%%%%%%%%%
\begin{figure}[t]
\centering
\includegraphics{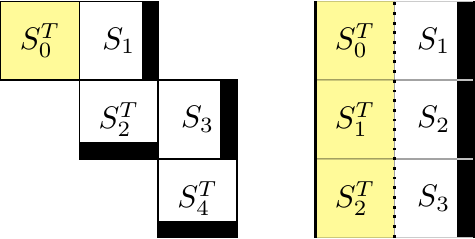}
\caption{Staircase code (left); corresponding zipper code (right).}
\label{fig:stczipper}
\end{figure}
%%%%%%%%%%%%%%%%%%%%%%%%%%%%%%%%%%%%%%%%%%%%%%%%%%%%%%%

\section{Examples}
\label{sec:examples}

This section gives some examples of related spatially-coupled codes that
can be described as zipper codes. Due to space constraints, only a few
codes will be described in detail, but other codes that can be described as
zipper codes include those in \cite{baggen,coe,northcott,humblet,g7093,montorsi}.

\subsection{Staircase Codes}

Staircase codes \cite{smith} are characterized by having an infinite
sequence of matrices of size $m\times m$: $S_0,S_1,\ldots$ such that, for
each $i \in \mathbb{N}$, every row of $\begin{bmatrix}S_{i}^T & S_{i+1}\end{bmatrix}$
is a codeword of a constituent code $\mathcal{C}$ of length $2m$ admitting
a systematic encoder, where
$S_{i}^T$ is the transpose of $S_{i}$.  The initial block $S_0$ is all zero
(and not transmitted).  The codes are so named because the sequence of
matrices form a staircase-like pattern when arranged as shown in
Fig.~\ref{fig:stczipper} (left), in which each row and each column must be
a codeword of $\mathcal{C}$, and where the dark-filled regions hold
parity symbols.

Staircase codes correspond to zipper codes with a fixed
constituent code $\mathcal{C}$ of length $2m$ admitting a systematic
encoder, and a zipping pair with $A_i = [m]$
for all $i\in \mathbb{N}$, i.e., the virtual positions comprise the
first $m$ positions in each row.
The interleaver map $\phi$ is $m$-periodic and defined to perform
a transposition operation, i.e.,
\[ \phi(mi+r,j)=\left(m(i-1)+j,m+r\right), \text{ for } r\in[m].
\]
The resulting buffer then forms the pattern shown in
Fig.~\ref{fig:stczipper} (right).

\subsection{Braided Block Codes}
\label{sec:braidedblockcodes}

\begin{figure}[t]
\centering
\includegraphics{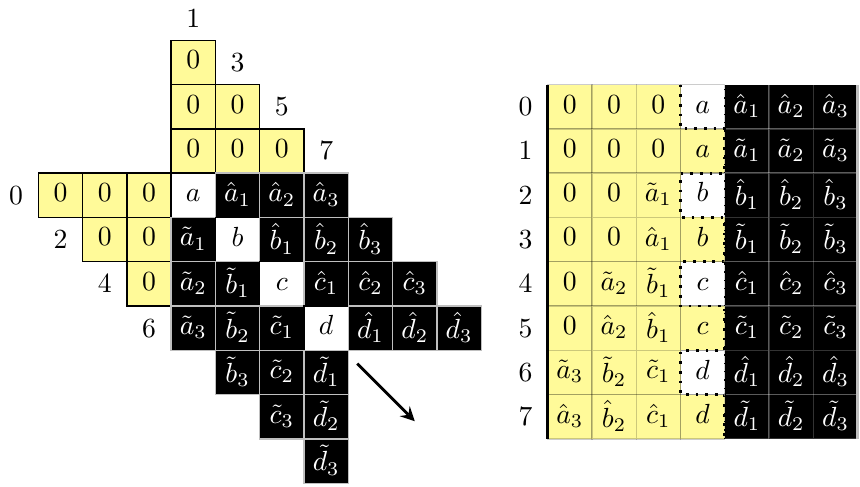}
\caption{Tightly braided code with $(7,4)$ Hamming constituent code (left);
corresponding zipper code (right), with numbers indexing constituent
codewords.}
\label{fig:bbczipper}
\end{figure}

Braided block codes \cite{feltstrom-bbc} are a type of
convolutional code whose codewords can be represented
as a subarray of an infinite two dimensional array
constrained by two interacting constituent
codes, one providing constraints on rows and the other
providing constraints on columns.
An example of a codeword from a rate $1/7$ tightly braided block code
from \cite{feltstrom-bbc} is shown in Fig.~\ref{fig:bbczipper} (left).
In this example, each row and each column in the diagram must
form a codeword of the binary $(7,4)$ Hamming code.  The information
positions are labelled $a,b,c,\ldots$ and the positions of
parity symbols are shown with dark-filled tiles. 

This tightly braided block code  corresponds to a zipper
code with fixed $(7,4)$ Hamming constituent code.  The zipping
pair is defined so that $A_i = [3]$ when $i$ is even,
and $A_i = [4]$ when $i$ is odd.
The interleaver map is given as
\[
\phi(i,j)=\begin{cases}
(i+2j-5,6-j) & \text{for } i \text{ even},\\
(i-2j-3,4+j) & \text{for } i \text{ odd, } j\neq 3,\\
(i-1,3) & \text{for } i \text{ odd, } j=3,
\end{cases}
\]
where the last case corresponds to copying an information
symbol from the row above.  In this example, only even-numbered
rows contain an information symbol.  The name ``zipper code''
derives from the buffer pattern formed in this example.

%%%%%%%%%%% FIGURE 3 %%%%%%%%%%%%%%%%%%%%%%%%%
%%%%%%%%%%%%%%%%%%%%%%%%%%%%%%%%%%%%%%%%%%%%%%%

\subsection{Diagonal Zipper Codes}
\label{sec:diagzipper}
In this subsection, we introduce two interleaver maps that couple the bits of a
zipper code in a regular (``hardware-friendly'') fashion.  These interleaver
maps generalize those of staircase codes.
\subsubsection{Tiled Diagonal Zipper Codes}

\begin{figure}[t]
\centering
\includegraphics{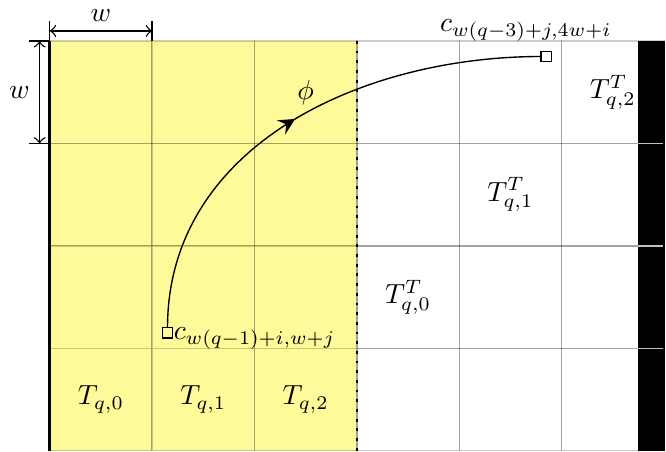}
\caption{Tiled diagonal zipper code with $L=3$, tile size $w\times w$, and
interleaver map as described in \eqref{eq:phitileddiagonal}.}
\label{fig:tiledzipper}
\end{figure}

Tiled diagonal zipper codes are a generalization of staircase codes that
are defined defined in terms of $w \times w$ ``tiles'' of symbols, a fixed
constituent code $\mathcal{C}$ with length $2Lw$ admitting a systematic
encoder, and a zipping pair with
$A_i = [Lw]$ (so that virtual positions comprise the first $Lw$ positions
in each buffer row), where the parameter $L$ is a positive integer.  The
interleaver map is defined so that each $w \times w$ tile within the
virtual buffer is the transpose of some $w \times w$ tile within the real
buffer, as illustrated in Fig.~\ref{fig:tiledzipper}.

To specify the interleaver map precisely, we must introduce a coordinate
system for tiles.  For a buffer $\boldsymbol{c}$, for any $s \in [2L]$ and for any $q
\in \mathbb{N}$, let
\[
T_{q,s}=\left[\begin{array}{ccc}
c_{wq,ws}&  \cdots &c_{wq,ws+w-1}\\
\vdots & \ddots & \vdots\\
c_{wq+w-1,ws}& \cdots &c_{wq+w-1,ws+w-1}
\end{array}\right].
\]
The $w \times w$ matrix $T_{q,s}$ is a tile located within the buffer whose
top left entry has position $(wq,ws)$.  When $s < L$, such a tile comprises
only virtual symbols, and therefore it is called a virtual tile; otherwise
it is a real tile.  For fixed $q$, the set of tiles $\{T_{q,s} \colon s
\in [2L] \}$ are said to form the $q$th \emph{tile row}.

For any $s \in [L]$ (indexing a virtual tile), we would
like to have $T_{q,s} = T_{q-s-1,L+s}^T$.  Thus the first virtual tile in
tile row $q$ is the transpose of the first real tile located one tile row
earlier, the second virtual tile in tile row $q$ is the transpose of the
second real tile located two tile rows earlier, and so on.  The interleaver
map that accomplishes this task is given as
\begin{equation}
\phi(wq+i,ws+j)=(w(q-s-1)+j,w(L+s)+i).
\label{eq:phitileddiagonal}
\end{equation}
Fig.~\ref{fig:tiledzipper} shows an example of a tiled diagonal zipper code
with $L=3$.

%%%%%%%%%%% FIGURE 4 %%%%%%%%%%%%%%%%%%%%%%%%%
%%%%%%%%%%%%%%%%%%%%%%%%%%%%%%%%%%%%%%%%%%%%%%

When $w=1$ (the case of $1\times 1$ tiles) tiled diagonal zipper codes
recover the interleaving pattern of the continuously interleaved BCH
(CI-BCH) codes described in \cite{coe}. When $L=1$, tiled diagonal zipper
codes are staircase codes.

\subsubsection{Delayed Diagonal Zipper Codes}

\begin{figure}[t]
\centering
\includegraphics{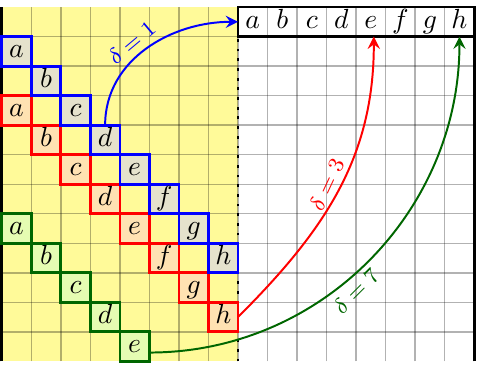}
\caption{Delayed diagonal zipper code with $m=8$ and various delay values~$\delta$.}
\label{fig:diagzipper}
\end{figure}

Delayed diagonal zipper codes are variants of tiled diagonal zipper codes
with $w=1$ and an added ``delay'' in the interleaver map. Specifically, the
interleaver map is given by
\[
\phi(i,j)=(i-j-\delta,j+m),
\]
where the positive integer $\delta$ is the delay parameter.  When
$\delta=1$, the interleaver map is identical to the interleaver map
(\ref{eq:phitileddiagonal}) with $w=1$.  Fig.~\ref{fig:diagzipper}
illustrates the buffer of a delayed diagonal zipper code with $m=8$.
As will show in Section~\ref{sec:stallpatternanalysis}, the introduction of
delay in the interleaver map reduces the multiplicity of minimal stall
patterns.  Delayed diagonal zipper codes can be generalized to the case of $w
\times w$ tiles, with $w > 1$, in the obvious way.

%%%%%%%%%%% FIGURE 5 %%%%%%%%%%%%%%%%%%%%%%%%%
%%%%%%%%%%%%%%%%%%%%%%%%%%%%%%%%%%%%%%%%%%%%%%

\section{Design Examples}
\label{sec:designexamples}

In this section, we present software simulation results
for several tiled diagonal and delayed diagonal zipper
code design examples.

\subsection{Simulation Setup}

We simulate transmission of zipper codewords over a binary symmetric
channel with crossover probability $p$.  This corresponds to binary
antipodal signalling over an additive white Gaussian noise channel
(Bi-AWGN) with hard-decision detection, but it may also model other
communications scenarios that give rise to independent and symmetrically
distributed bit errors.  The \emph{Shannon limit} at rate $R$ for such a
channel is the largest crossover probability $p$ for which it is
theoretically possible to communicate at rate $R$ (bit/channel use) with
arbitrarily small probability of error.  For the Bi-AWGN, this value of $p$
is achieved at some particular signal-to-noise ratio (SNR).   A code of
finite length will achieve sufficiently small error probability only at
some crossover probability smaller than the Shannon limit, corresponding to
a larger Bi-AWGN SNR.  The ``gap to the Shannon limit'' is then defined as the
difference (in decibels) between the Bi-AWGN SNR corresponding to the
Shannon limit and the Bi-AWGN SNR at which the code achieves sufficiently
small error probability.

By ``sufficiently small error probability'' we mean a post-correction
(post-FEC) bit error rate (BER) of $10^{-15}$ or smaller.  As we cannot reliably
measure such low error rates in any reasonable amount of time using
software simulations, we use a least-squares fit linear extrapolation (on
the log-log scale) of our measured error rates, in order to estimate the
BSC crossover probability $p^*$ at which the post-FEC BER achieves
$10^{-15}$.

In all of our design examples, we use shortened $t$-error-correcting $(n,k)$
BCH constituent codes, where $n=2m$.  We made no attempt to optimize the
positions at which the code is shortened.  The virtual buffer and the real
buffer both have width $m$.  We use a causal, periodic, bijective interleaver
map that implements a diagonal or delayed diagonal zipper code.

Decoding is performed using exhaustive decoding, with at most five rounds
per window of size $M$ rows (thus containing $Mm$ real symbols).  We use
chunk decoding with chunk size $m$, fresh/stale flags, and periodic
truncation with transmission length $J=995m$ and $\tau=5m$.

\subsection{Varying Decoding Window Sizes}

\begin{figure}[t] \centering
\includegraphics{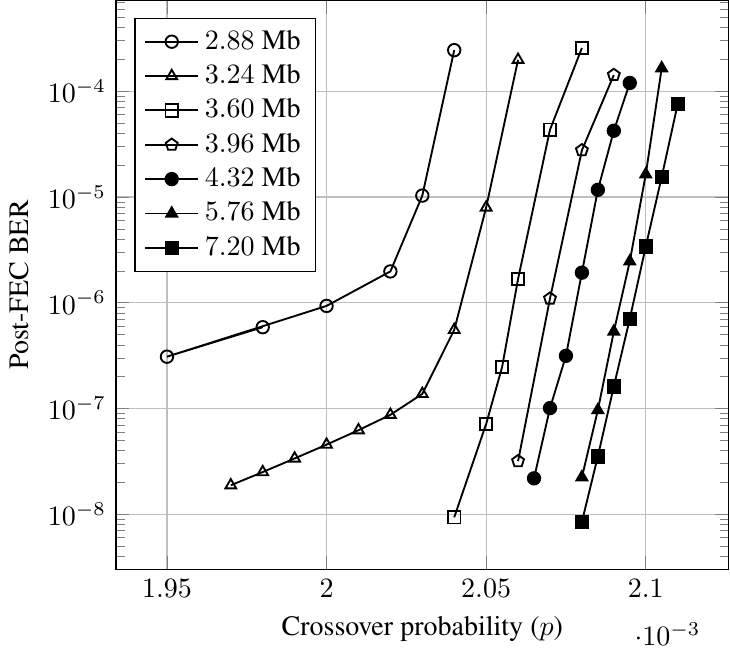}
\caption{Simulation results
of rate $0.97$ tiled diagonal zipper codes ($w=1$, $t=3$ constituent code) with different decoding
window size.} \label{fig:simbufsize} \end{figure}

We simulated a rate $0.97$ tiled diagonal zipper code with $w=1$ and
$m=1200$, (corresponding to a CI-BCH code \cite{coe}) having a
$(2400,2364,t=3)$ shortened BCH constituent code, over varying decoding
window sizes. As shown in Fig.~\ref{fig:simbufsize}, we observe an
improvement in the decoding performance as well as disappearing error flare
as we increase the decoding window sizes. This is due to the fact that
larger decoding windows provide each symbol with a larger number of
decoding rounds. However, as expected, we see a diminishing return as the
window size increases, with only slight improvement when increasing the
decoding window size from $4m^2 = 5.76$~Mb to $5m^2 = 7.20$~Mb.  Simulation
results for a variety of different tiled diagonal zipper codes indicate
that choosing the window size near $5m^2$ gives nearly best possible
decoding performance.   This choice of window size is
a convenient heuristic, akin to the well-known rule-of-thumb
in convolutional decoding that suggests that
a \emph{truncation depth} of about five times the
code's \emph{constraint length} suffices to provide nearly
optimal decoding performance under Viterbi decoding (see, e.g.,
\cite{forney1974}).  In a tiled diagonal zipper code
with $w=1$, the parameter $m^2$ plays the role of constraint
length, since output symbols depend on input symbols located
as far as $m$ rows earlier (and each row contains $m$ real symbols).

%% FIGURE 6 %%%%%%%%%%%%%%%%%%%%%%%%%%%%%%%%%%%%%%%%%%%%%%%%%%%%
%%%%%%%%%%%%%%%%%%%%%%%%%%%%%%%%%%%%%%%%%%%%%%%%%%%%%%%%%%%%%%%%
%% FIGURE 7
\begin{figure}[t] \centering
\includegraphics{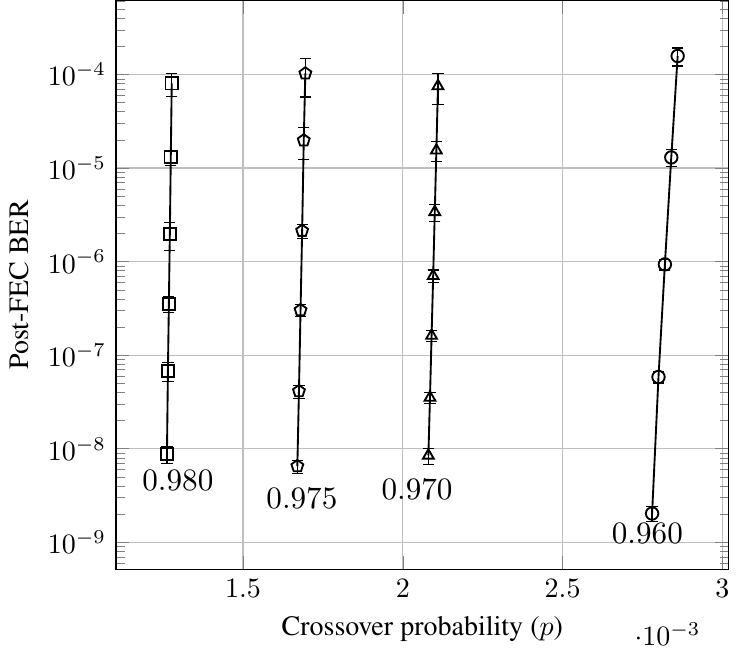}
\caption{Simulation results of tiled diagonal zipper codes ($w=1$, $t=3$ constituent code) over different
code rates. Error bars are located one standard deviation from the mean.}
\label{fig:diagrate} \end{figure}
%%%%%%%%%%%%%%%%%%%%%%%%%%%%%%%%%%%%%%%%%%%%%%%%%%%%%%%%%%%%%%%%
%%% FIGURE 8
\begin{figure}[t] \centering
\includegraphics{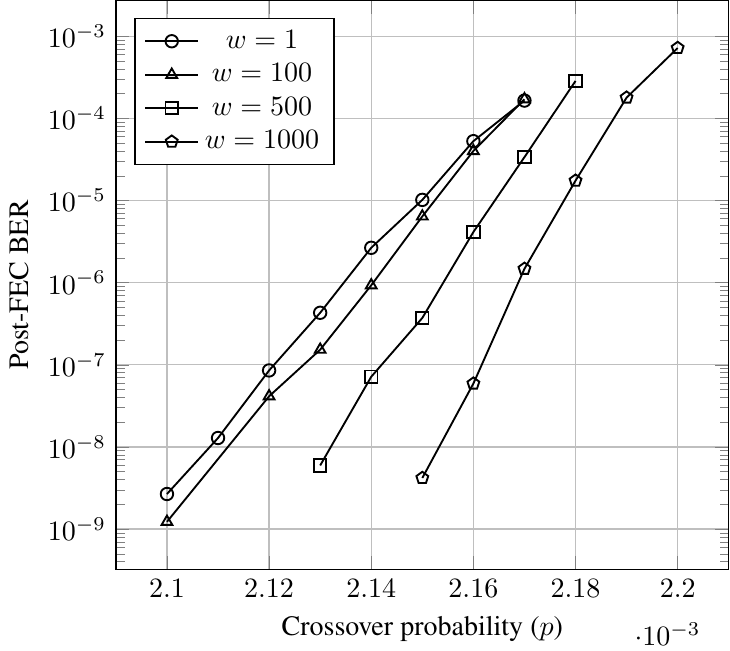}
\caption{Simulation results of rate $0.967$ tiled diagonal zipper codes ($t=3$ constituent code) with
different tile sizes.} \label{fig:simtileddiag} \end{figure}
%%%%%%%%%%%%%%%%%%%%%%%%%%%%%%%%%%%%%%%%%%%%%%%%%%%%%%%%%%%%%%%%
%%% FIGURE 9
\begin{figure}[t] \centering
\includegraphics{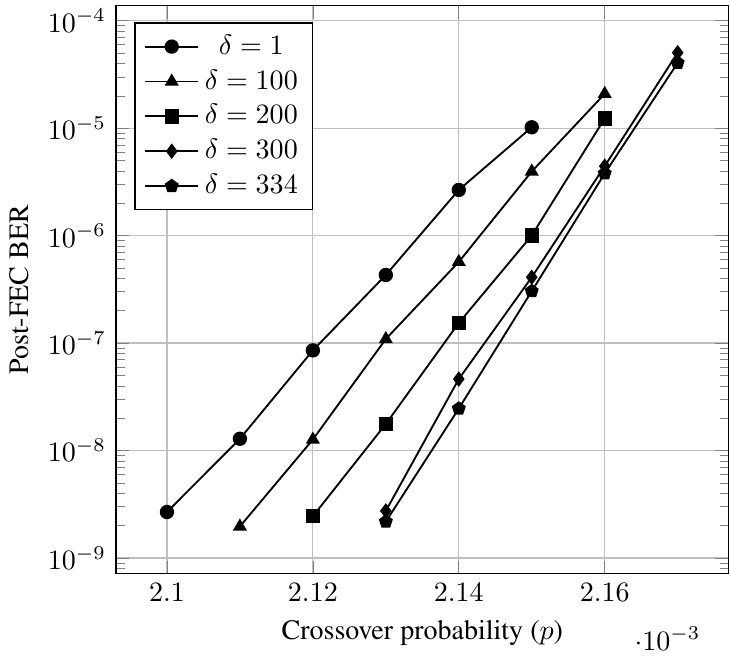}
\caption{Simulation results of rate $0.967$ delayed diagonal zipper codes ($t=3$ constituent code) with
different delay parameters.} \label{fig:simdelayeddiag} \end{figure}
%%%%%%%%%%%%%%%%%%%%%%%%%%%%%%%%%%%%%%%%%%%%%%%%%%%%%%%%%%%%%%%%

\subsection{Different Code Rates}
\label{sec:diffcoderates}

As shown in Fig.~\ref{fig:diagrate}, we simulated the decoding of tiled
diagonal zipper codes with tile size $w=1$ having different code rates, as
summarized in Table~\ref{tab:sim1}.  The decoding window (DW) size was set to
$5m^2$ in all cases.  We see that the gap to the Shannon limit decreases as the rate
increases, which is consistent with the theoretical analysis of
\cite{jian}.  

\begin{table}[b]
\centering
\caption{Simulation parameters, thresholds, and gaps to the Shannon limit.}
\setlength{\tabcolsep}{2pt}
\begin{tabular}{c|c|c|c|c|c}
Rate & $m$ & $(n,\,k,\,t)$ & DW Size & $p^*$ & Gap (dB)\\\hline
$0.960$ & $825$  & $(1650,1617,3)$ & $3.4$ Mb & $2.68\cdot10^{-3}$ & $0.503$\\
$0.970$ & $1200$ & $(2400,2364,3)$ & $7.2$ Mb & $2.03\cdot10^{-3}$ & $0.412$\\
$0.975$ & $1440$ & $(2880,2844,3)$ & $10.4$ Mb & $1.63\cdot10^{-3}$ & $0.398$\\
$0.980$ & $1800$ & $(3600,3564,3)$ & $16.2$ Mb & $1.24\cdot10^{-3}$ & $0.393$\\
    \end{tabular}
\label{tab:sim1}
\end{table}

\subsection{Tiled Diagonal and Delayed Diagonal Zipper Codes}

We simulated rate $0.967$ codes with $m=1000$, based on a $(2000,1967,t=3)$
shortened BCH constituent code, with $5$~Mb decoding window size.  Various
tiled and delayed diagonal zipper codes having a variety of tile sizes $w$
and delay parameter $\delta$, as summarized in
Table~\ref{tab:tileddelayeddiag}, were simulated.  Performance curves for
the tiled diagonal codes are shown in Fig.~\ref{fig:simtileddiag} and those
for the delayed diagonal codes are shown in Fig.~\ref{fig:simdelayeddiag}.

In both cases, increasing $w$ or $\delta$ helps improve the decoding with
an improvement to the gap to the Shannon limit of $0.028 \sim 0.039$ dB when
compared to the $\delta=w=1$ case.  A stall pattern analysis for tiled
diagonal and delayed diagonal zipper codes, which may provide theoretical
justification for these performance improvements, is given in
Sec.~\ref{sec:stallpatternanalysis}.

%%%%%%%%%%% FIGURE 10 %%%%%%%%%%%%%%%%%%%%%%
\begin{figure}[t] \centering
\includegraphics{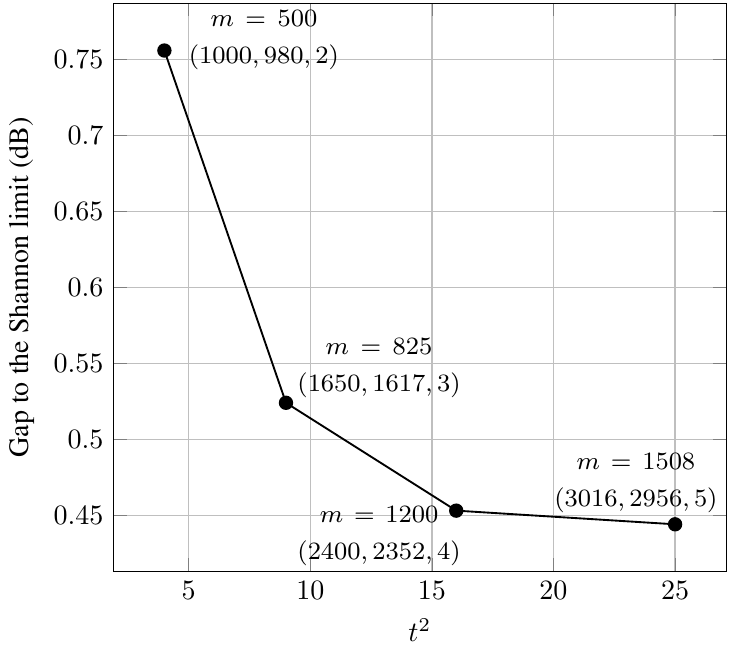}
\caption{Gap to the Shannon limit for different decoding radius of a rate $0.96$ delayed
diagonal zipper code with $\delta=1$. The values of $m$ and constituent code
parameters $(n,k,t)$ are shown next to each point. In all cases, we have the decoding
window size to be $5m^2$ bits.} \label{fig:perfvscomp}
\end{figure}
%%%%%%%%%%%%%%%%%%%%%%%%%%%%%%%%%%%%%%%%%%%%%%

\begin{table}[b]
  \centering
  \caption{Gap to the Shannon limit at $10^{-15}$ post-FEC BER for tiled diagonal and delayed diagonal zipper codes.}
  \begin{tabular}{c|c|c|c}
    Type & $w$ or $\delta$ & $p^*$ & Gap (dB)\\ \hline
    tiled/delayed & $1$ & $2.015\cdot 10^{-3}$ & $0.536$\\
    tiled & $100$ & $2.036\cdot 10^{-3}$ & $0.526$\\
    tiled & $500$ & $2.060\cdot 10^{-3}$ & $0.514$\\
    tiled & $1000$ & $2.099\cdot 10^{-3}$ & $0.497$\\\hline
    delayed & $100$ & $2.035\cdot 10^{-3}$ & $0.526$\\
    delayed & $200$ & $2.048\cdot 10^{-3}$ & $0.520$\\
    delayed & $300$ & $2.071\cdot 10^{-3}$ & $0.509$\\
    delayed & $334$ & $2.073\cdot 10^{-3}$ & $0.508$\\
  \end{tabular}
  \label{tab:tileddelayeddiag}
\end{table}

\subsection{Performance versus Complexity}

Let us define the \emph{performance} of a zipper code as its gap to the Shannon limit at
$10^{-15}$ output bit error rate, and its \emph{complexity} as the squared decoding
radius $t^2$ of its constituent code.  This complexity
measure is chosen in view of the  well-known fact that
the power consumption of a
Berlekamp-Massey-based BCH decoder circuits grows as
$\mathcal{O}(t^2)$ \cite{liu}.
Fig.~\ref{fig:perfvscomp} plots performance versus complexity for
various rate $0.96$ diagonal zipper codes.
As expected,
the gap to the Shannon limit is
reduced by increasing the complexity of the constituent code,
but with diminishing returns.

\section{Stall Pattern Analysis}
\label{sec:stallpatternanalysis}

This section will characterize stall patterns of zipper codes.
For simplicity, throughout this
section we will be focusing on zipper codes whose interleaver maps are
bijective.

\subsection{Graph Representation of Zipper Codes}

Due to the spatially-coupled structure of zipper codes, it can be helpful to
describe them as codes defined on graphs.  Following the notation of
Sec.~\ref{sec:structure}, we consider a zipper code corresponding
to constituent code sequence $\mathcal{C}_0, \mathcal{C}_1, \ldots$,
with respective information sets $I_0, I_1, \ldots$.  For each
$i \in \mathbb{N}$, the virtual symbols in the $i$th row of a
buffer are indexed by the set $A_i \subseteq I_i$.
Fix a bijective interleaver map $\phi = (\phi_1, \phi_2)$.
Let
\[
\phi_1(A_i)=\left\{\phi_1(i,j) \colon j \in A_i\right\} \cap \mathbb{N}
\]
be the set of row indices from which the virtual symbols in row $i$
are copied (excluding any negative row indices).

Define a graph $G = (V,E)$ whose vertex set $V = \mathbb{N}$ is
the set of natural numbers, and whose edge set
\[
E = \{ \{ i,j \}: i \in \mathbb{N}, j \in \phi_1(A_i) \}.
\]
The vertices then correspond to constituent codes,
with an edge joining two vertices if the corresponding codes
have a symbol in common.    Parallel edges are permitted,
thus we interpret the edge set as a
multiset (a set in which elements may have multiplicity greater than one).
The number of edges between two vertices is then equal to
the number of symbols that the corresponding codes have in common.
Allowing for parallel edges, the degree of vertex $i$ is
equal to $n_i$, the block length of $\mathcal{C}_i$, unless some
virtual symbols in row $i$ are copied from rows with negative
row index, in which case the corresponding edges are missing.

\begin{figure}[t]\centering
\includegraphics{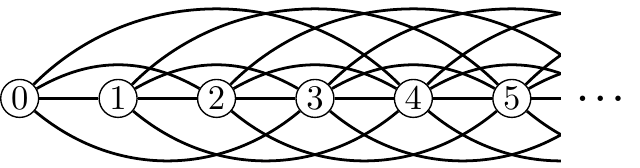}
\caption{Graph representation of a tiled diagonal zipper code with $m=4$, $w=1$.}
\label{fig:diag-graph}\end{figure}

\begin{example}
Recall that the interleaver map of a tiled
diagonal zipper code with $w=1$ is given by $\phi(i,j)=(i-j-1,m+j)$.
For each $i \in \mathbb{N}$, we have
$\phi_1(A_i) = \{ i-1, i-2, \ldots, i-m \} \cap \mathbb{N}$.  Thus
vertex $i$ will have neighbors
\[
\mathcal{N}(i) = \{ i \pm 1, i \pm 2, \ldots, i \pm m \} \cap \mathbb{N}.
\]
Vertices $m, m+1, m+2, \ldots$ therefore have degree $n=2m$.
The graph of a tiled diagonal zipper code with $m=4$, $w=1$ is shown
in Fig.~\ref{fig:diag-graph}.
\label{ex:diagzipper}
\end{example}

%%%%%%%%%%%%% FIGURE 11 %%%%%%%%%%%%%%%%%
%%%%%%%%%%%%%%%%%%%%%%%%%%%%%%%%%%%%%%%%%

Graph representations for zipper codes in which symbols are copied more than
once (i.e., when the interleaver map is non-bijective) can be defined by
appealing to hypergraphs (in which hyperedges comprise more than two vertices),
to a bipartite representation such as a factor graph, or by introducing
additional constituent repetition-code constraints.  We will not pursue such
representations in this paper.

\subsection{Error and Stall Patterns}

Recall that $B$ is a set containing the positions of all real symbols in a
buffer.  An \textit{error pattern} $S \subseteq B $ is any nonempty subset of $B$
containing the positions of erroneous symbols in the real part of a received
buffer.  For any error pattern $S$, we define
\[
S^* = S \cup \bigcup_{s \in S} \phi^{-1}(s)
\]
to be the complete set of positions of errors and their duplicates in the
virtual set, and we define $\pi(S^*)=\{i:(i,j)\in S^*\}$ to be the set of
\textit{affected rows}.

For simplicity, throughout this section we assume that the constituent code is
identical for all rows, i.e., $\mathcal{C}_i=\mathcal{C}$ with decoding radius
$t_i=t$ for all $i\in\mathbb{N}$. In addition, we assume that we use a
genie-aided, miscorrection-free constituent decoder.  Such a decoder is able to
correct up to $t$ errors in a row, but always fails to decode (never
miscorrecting) when the number of errors in a row exceeds $t$.  We assume that
this genie-aided decoder visits rows in a decoding window as many times as is
needed, reducing the error pattern size until no further correction is
possible. Should some errors remain after this decoding procedure, we call the
remaining errors a \emph{stall pattern}. In a stall pattern, we must have at
least $t+1$ errors in every remaining affected row.

We will consider only strictly causal interleaver maps that induce at most
one shared symbol between any two distinct rows of a zipper code.
This means that the resulting graph is  \emph{simple}, i.e.,
it contains no parallel edges or self-loops.  Each edge of
a such a simple graph corresponds to one codeword symbol.
We refer to such an interleaver map as \emph{scattering}.

\subsection{Decoding on a Graph}

Assuming a scattering interleaver map,
an error pattern $S$ can be represented as a subgraph
of the graph representing the zipper code.
If the zipper code has
graph representation $G=(V,E)$, the graph representation of $S$ is
$G_S=(V_S,E_S)$, where $V_S= \pi(S^*) \subseteq V$ (i.e., the
vertex set of $S$ contains vertices corresponding to the rows
affected by $S$) and
$E_S =  \{ \{ i,j \}: (i,k) \in S, j \in \phi_1^{-1}(i,k) \}$
(i.e.,  the edges in $E_S$ correspond to symbols in error).
From this construction, the number of edges in $G_S$ is exactly the cardinality of $S$.
We call $|S|=|E_S|$ the \textit{size} of the error pattern $S$.

Suppose that the constituent code $\mathcal{C}$ can correct up to $t$ errors.
The action of the genie-aided decoder can then be described
as an edge-peeling process in the error pattern graph $G_S=(V_S,E_S)$:
\begin{enumerate}
    \item For each vertex $v \in V_S$ with $\deg(v) \leq t$,
    remove $v$ from $V_S$ as well as all edges in $E_s$ incident on $v$.
    \item Repeat step 1 until either the graph is empty or all
remaining vertices have degree $t+1$ or higher.
\end{enumerate}
This procedure will terminate either in an empty graph
(successful decoding),
or in a $(t+1)$-core of $G_S$---the largest induced subgraph
of $G_S$ with the property that all vertices have degree at least $t+1$.
Sometimes, though not always,
the $(t+1)$-core will form a \emph{$(t+2)$-clique}, i.e., a 
subgraph of $G_S$ comprising $t+2$ vertices all of which
are neighbors (a complete subgraph).
Such a pattern is uncorrectable by the genie-aided decoder.
More generally, we call $G_S$ a \textit{stall pattern}
if each vertex in $G_S$ has degree at least $t+1$, making it
uncorrectable.

The following example illustrates error and stall patterns
in both array and graph representations.

\begin{figure}[t]
    \centering
    \includegraphics{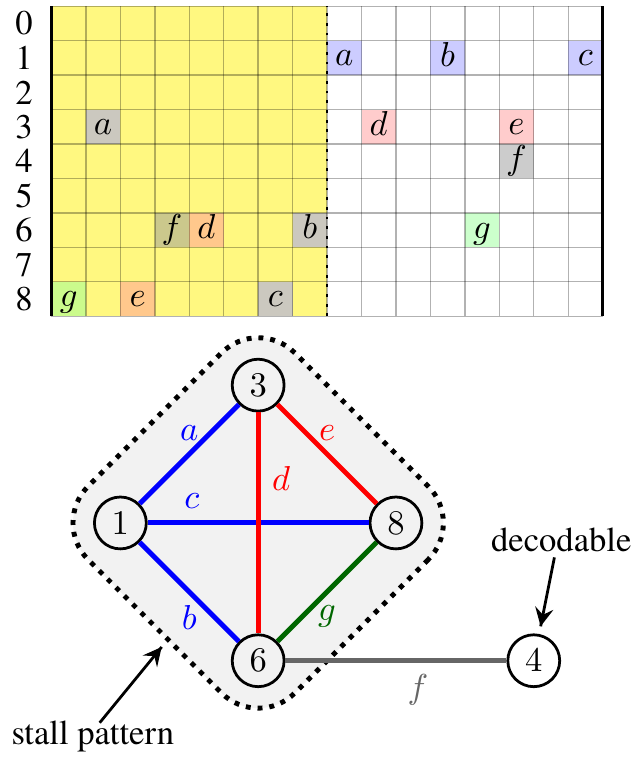}
    \caption{Example of an error pattern with underlying stall pattern in both array (top) and graph (bottom) forms. In this example, the constituent code is double-error-correcting ($t=2$).}
    \label{fig:stallpat}
\end{figure}

\begin{example}
Consider a zipper code with a constituent code $\mathcal{C}$ capable
of correcting $t=2$ errors.
Suppose that we receive an error pattern with seven errors as shown in Figure
\ref{fig:stallpat}. 
Since rows $1,3,6,$ and $8$ have at least $3$ errors each, we cannot
correct them. However, we can still correct row $4$, which has only one error.
Thus we will remove the error from row $4$ and its corresponding entry in row
$6$, or equivalently, we remove vertex $4$ and its incident edges.
However, the remaining errors cannot be corrected and so those errors form a
stall pattern. The $4$-clique formed by vertices $1, 3, 6,$ and $8$ is a
stall pattern and it is also the $3$-core of the error pattern graph
from which we started.
\end{example}

%%%%%%%%%%%%% FIGURE 12 %%%%%%%%%%%%%%%%%
%%%%%%%%%%%%%%%%%%%%%%%%%%%%%%%%%%%%%%%%%

\subsection{Properties of Stall Patterns of Zipper Codes}

The following theorem bounds the size of a stall pattern
in a zipper code with a bijective and scattering interleaver map.
\begin{theorem}
A stall pattern $S$ for a zipper code with a $t$-error-correcting
constituent code
and having a bijective and scattering interleaver map satisfies
$|S| \geq \frac{1}{2}(t+1)(t+2)$.
\label{thm:minstall}
\end{theorem}
\begin{IEEEproof}
Let $G_S=(V_S,E_S)$ be the graph representation of stall pattern $S$, and let
$v$ be any vertex in $V_S$. Then, since $G_S$ is a stall pattern, $\deg(v) \geq
t+1$. Suppose that $\mathcal{N}_S(v)=\{v_1,v_2,\ldots,v_{\deg{v}} \}$ are the
neighbors of $v$ in $G_S$.  Then, since each $v_i$ is a vertex in a stall
pattern, $\deg(v_i) \geq t+1$.  Hence, there must be at least
$ \deg(v) +1\geq t+2$ vertices in $G_S$ with each vertex having degree $t+1$ or higher.  We now apply the
handshaking lemma of graph theory \cite[Prop.~1.3.3.]{west_2000},
which 
states that for every (finite) graph $(V,E)$
we have $\sum_{v\in V}\deg(v)=2|E|$, since each edge of the graph is counted
exactly
twice when summing over vertex degrees.
Thus it follows that
\[
|S|=|E_S|=\frac{1}{2}\sum_{v\in V_S}\deg(v) \geq \frac{1}{2}(t+1)(t+2).
\]
\end{IEEEproof}

It is worth noting that $\frac{1}{2}(t+1)(t+2)$ is precisely the number of
edges of a complete graph with $t+2$ vertices. In fact, we will now
show that the existence of a $(t+2)$-clique in the graph representation of the
zipper code is a necessary and sufficient condition for the existence of a
stall pattern of size $\frac{1}{2}(t+1)(t+2)$.
\begin{prop}
Let $G=(V,E)$ be the graph representation of a zipper code with a
$t$-error-correcting constituent code and a bijective and scattering
interleaver map.  Then the code has a stall pattern $S$ of size
$\frac{1}{2}(t+1)(t+2)$ if and only if $G$ contains a $(t+2)$-clique.
  \label{prop:smallstall}
\end{prop}
\begin{IEEEproof}
  $(\Leftarrow)$ Every $(t+2)$-clique corresponds to a stall pattern of
size $\frac{1}{2}(t+1)(t+2)$.\\
$(\Rightarrow)$ Suppose by way of contradiction that a stall pattern $S$
of size $\frac{1}{2}(t+1)(t+2)$ has a graph $G_S$ with more than $t+2$ vertices.
The average
degree of the vertices is then less than $2|S|/(t+2)=t+1$,
which implies implies that $V_S$
contains at least one vertex of degree $t$ or lower, contradicting
the fact that $S$ is a stall pattern. 
On the other hand, if $G_S$ contains fewer than $t+2$ vertices,
then there exists a vertex in $V_S$ with degree greater than
$t+1$, which is impossible in a simple graph.
\end{IEEEproof}

\begin{figure}[t]
\centering
\includegraphics{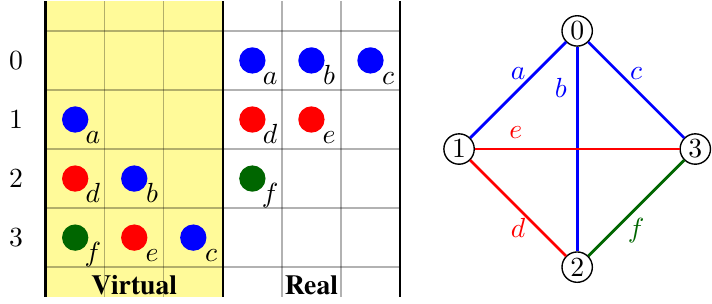}
\caption{Graph representation of a stall pattern of size $6$ of tiled diagonal
zipper code, $w=1$, $m=3$, $t=2$.}
\label{fig:diagzipperstall}
\end{figure}

An example of an interleaver map that yields a stall pattern of size
$\frac{1}{2}(t+1)(t+2)$ is the tiled diagonal zipper code with tile size $1$
(or delayed diagonal with delay $1$). Fig.~\ref{fig:diagzipperstall} shows an
example of a tiled diagonal zipper code with $w=1$, $m=3$, $t=2$.

%%%%%%%%%%%%% FIGURE 13 %%%%%%%%%%%%%%%%%
%%%%%%%%%%%%%%%%%%%%%%%%%%%%%%%%%%%%%%%%%

Theorem~\ref{thm:minstall} can be generalized to
strictly causal non-scattering
interleaver maps that allow constituent codes to have up to $b$
bits in common, so that the resulting graph representation has
up to $b$ parallel edges connecting any two vertices.
\begin{theorem}
For a zipper code with $t$-error-correcting constituent code having
a strictly causal interleaver map that allows up to $b$ parallel
edges between any pair of vertices in its graph, the number of errors
in a stall pattern is at least
\[
\frac{1}{2}\left( 1 + \left\lceil \frac{t+1}{b} \right\rceil \right)(t+1).
\]
\end{theorem}
\begin{IEEEproof}
Any vertex in a stall pattern graph
must have at least $\lceil (t+1)/b \rceil$ neighbors. The result
then  follows from the same argument as in the proof
of Theorem~\ref{thm:minstall}.
\end{IEEEproof}

\subsection{Error Floor Approximation}

The presence of stall patterns creates so-called ``error floors'' in the
performance curves of zipper codes.  We may approximate the location of the
error floor using a union bound technique similar that used in \cite{smith}. We
consider a decoding window of size $M\times m$, and denote the set of all stall
patterns in the decoding window to be $\mathcal{S}$. We determine the error
floor estimate by enumeration of $\mathcal{S}$, evaluating (an upper
bound on)
the probability that the particular error pattern arises
at the output of a binary symmetric channel with crossover probability
$p$.
This then gives
\begin{align}
\text{BER}_{\text{floor}} \leq \frac{1}{Mm}\sum_{S\in\mathcal{S}}|S|p^{|S|}.
\label{eq:errfloor}
\end{align}
Suppose that we could determine exactly the sizes of stall patterns that
can occur in a
decoding window. We could then rewrite  \eqref{eq:errfloor} as
\begin{align}
\text{BER}_{\text{floor}} \leq \frac{1}{Mm}\sum_{\ell\in\mathcal{L}}N_\ell p^{\ell},
\label{eq:errfloor2}
\end{align}
where $\mathcal{L}$ denotes the set of all stall-pattern sizes that can occur
in the decoding window of size $Mm$ and $N_\ell$ denotes the number of
occurrences of stall patterns of size $\ell$. Note that since we only consider
stall patterns that can fit in the decoding window, we have $\ell\leq Mm$.

The possible sizes and the number of occurrences of stall patterns of certain
size depend on the interleaver map. For example, it is possible to construct
stall patterns of size $\frac{1}{2}(t+1)(t+2)$ in tiled diagonal zipper codes
with tile size $1$, but not in staircase codes, whose smallest stall pattern
size is $(t+1)^2$ \cite{smith,holzbaur}.

Given a decoding window, a set $\mathcal{L}$ of possible stall pattern sizes
that can fit in the window, and crossover probability $p$, we call the stall
patterns of size $\ell^*\in\mathcal{L}$ to be \emph{dominant} if
$N_{\ell^*}p^{\ell^*}\geq N_\ell p^\ell$ for all $\ell\in\mathcal{L}$.  In
general, the minimum-sized stall patterns may not be dominant since it is
possible that the multiplicity of larger stall patterns causes a dominant
$N_\ell p^ \ell$ term. However, for sufficiently small  $p$, we can assume that
stall patterns of minimum size are dominant.  The error floor can be then
further approximated as the contribution of just the minimum-sized stall
patterns, i.e., the right hand side of \eqref{eq:errfloor2} is approximately
$N_{\ell^*}p^{\ell^*}$, where $\ell^*=\min\mathcal{L}$.

\subsection{Eliminating Small-Sized Stall Patterns of Diagonal Zipper Codes}

We will now describe a few strategies to eliminate stall patterns of size
$\frac{1}{2}(t+1)(t+2)$ in tiled and delayed diagonal zipper codes.

\subsubsection{Tiled Diagonal}

Our first observation is that
we can reduce the multiplicity stall patterns of size
$\frac{1}{2}(t+1)(t+2)$ by increasing the tile size. To see this, we will first
count the occurrences of stall patterns of size $\frac{1}{2}(t+1)(t+2)$.
Denote the tile size as $w$ and assume that $m=wL$, $M=wK$
($K>L$), and $L\geq t+1$. In order to construct a stall pattern of size
$\frac{1}{2}(t+1)(t+2)$, we first pick $t+1$ tiles from the same row. We
then select a row index at which to place the  errors. For each of the $t+1$
tiles, pick one column index at which to place an error. Those errors
will correspond to errors in $t+1$ other rows. The positions of
the remaining $\frac{1}{2}t(t+1)$ errors in those rows are then forced.
Thus, the number of stall patterns of size
$\frac{1}{2}(t+1)(t+2)$ is given by
\begin{align}
\sum_{s=t}^{L-1}\binom{s}{t}(K-s-1)w^{t+2}\approx \binom{L}{t+1}Kw^{t+2}.
\label{eq:tileddiagstall}
\end{align}
Having larger tile size will make the values for $K$ and $L$ lower assuming
that the decoding window size is fixed. This will in turn reduce the
multiplicity
of stall patterns of size $\frac{1}{2}(t+1)(t+2)$ given by
\eqref{eq:tileddiagstall}. Furthermore once $L\leq t$, it is impossible to
form a stall pattern of size $\frac{1}{2}(t+1)(t+2)$.

\subsubsection{Delayed Diagonal}

In delayed diagonal zipper codes, the occurrences and size of the
minimum-sized stall patterns depends on $\delta$. In particular, we will
show that having larger delay reduces the multiplicity of stall patterns of
size $\frac{1}{2}(t+1)(t+2)$.
\begin{prop}
For a delayed diagonal zipper code with $t$-error-correcting constituent
code and delay parameter $\delta$, there exists a stall pattern of size
$\frac{1}{2}(t+1)(t+2)$ if and only if $\delta\leq\frac{m-1}{t}$.
\end{prop}
\begin{IEEEproof}
Without loss of generality, let the first affected row index of the stall
pattern be zero.\\
$(\Leftarrow)$ We claim that we can form a stall pattern of size
$\frac{1}{2}(t+1)(t+2)$ whose set of row indices is given by
$I=\{0,\delta,2\delta,\ldots,(t+1)\delta\}$. To see this, first observe
that $(t+1)\delta=t\delta+\delta\leq m+\delta-1$.  The neighbors of row
$i$ in the full zipper code graph are then indexed by
$$\mathcal{N}(i)=\{i\pm(\delta+j):j\in[m]\}\cap\mathbb{N}.$$  Thus, $i+(t+1)\delta\leq\max(\mathcal{N}(i))$
and $\max\{0,i-(t+1)\delta\}\geq\min(\mathcal{N}(i))$, so $I\subseteq\mathcal{N}(i)$ for all $i\in
I$. It follows that the rows in $I$ are connected with each other and so it
is possible to form a $(t+2)$-clique from $I$.\\
$(\Rightarrow)$ Let $I=\{0,i_1,\ldots,i_{t+1}\}$ with
$0<i_i<\ldots<i_{t+1}$ be the row indices of a stall pattern of size
$\frac{1}{2}(t+1)(t+2)$. Then it must be true that
\begin{align*}
i_1 &\geq 0+\delta=\delta,\\
i_2 &\geq i_1+\delta \geq 2\delta,\\
 &\vdots\\
i_{t+1}&\geq i_t+\delta \geq (t+1)\delta.
\end{align*}
However, we also require that row $i_{t+1}$ to be a neighbor of the first
row, i.e., we require $i_{t+1}\leq m+\delta-1$. Hence, $(t+1)\delta\leq
m+\delta-1$, and rearranging yields $\delta\leq \frac{m-1}{t}$.
\end{IEEEproof}

Suppose that $\delta\leq \frac{m-1}{t}$. In order to construct a stall
pattern of size $\frac{1}{2}(t+1)(t+2)$ with the set of affected row index
$I=\{i_0=0,i_1,\ldots,i_{t+1}\}$, the following constraints must be
satisfied:
\begin{itemize}
\item Errors in row $0$ must affect row $i_{t+1}$, i.e.,
$i_{t+1}\leq \delta+m-1$.
\item For $j=1,\ldots,t+1$, $i_j-i_{j-1}\geq \delta$.
\end{itemize}
Thus, the indices $i_1,\ldots,i_{t+1}$ must be taken from $\{\delta,
\delta+1,\ldots, \delta+m-1\}$ and any two indices must differ by at least
$\delta$. 
The number of possible combinations of such indices is given by
\[
\binom{m-t(\delta-1)}{t+1}=\binom{m-t\delta+t}{t+1}.
\]
In a decoding window with $M$ rows, the number of possible combinations is
therefore at most
\begin{align*}
M\binom{m-t\delta+t}{t+1}.
\label{eq:delayeddiagstall}
\end{align*}

\begin{figure}[t]
\centering
\includegraphics{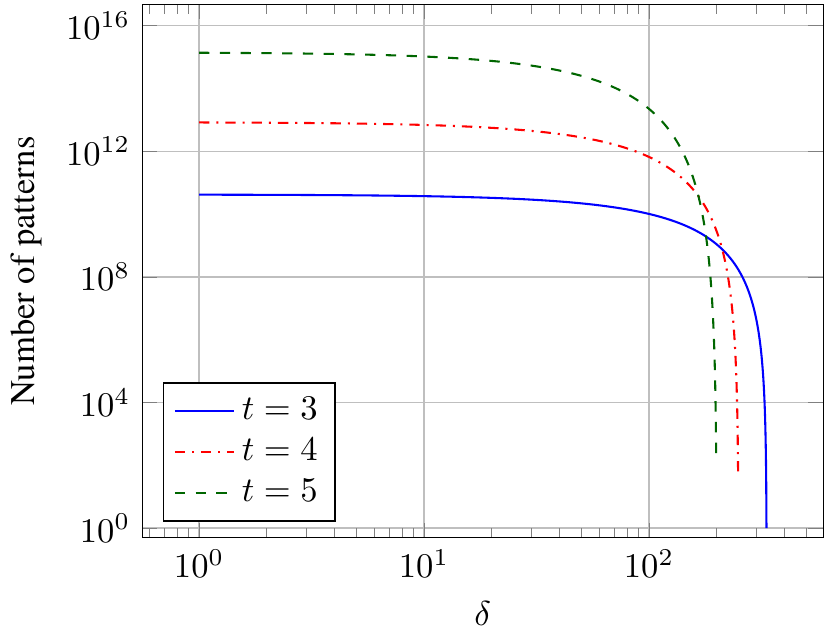}
\caption{Number of stall patterns of size $\frac{1}{2}(t+1)(t+2)$
involving the first row of the decoding window
in delayed diagonal zipper codes with $m=1000$ and varying $t,\delta$.}
\label{fig:deldiag_num}
\end{figure}

\begin{example}
Figure \ref{fig:deldiag_num} shows the number of stall patterns of size
$\frac{1}{2}(t+1)(t+2)$ involving the first row of
the decoding window in delayed diagonal zipper codes for $m=1000$ and
$t=3,4,5$. Observe that when $\delta=1$ and $t=3$, there are
$$\binom{1000}{4}\approx 4.14\times 10^{10}$$ possible configurations
involving 
the first row. However, there is only one possible configuration for
$\delta=333$ and stall patterns of size $\frac{1}{2}(t+1)(t+2)$ do not
exist for $\delta\geq 334$.
\end{example}

%%%%%%%%%%%%% FIGURE 14 %%%%%%%%%%%%%%%%%
%%%%%%%%%%%%%%%%%%%%%%%%%%%%%%%%%%%%%%%%%

\section{Conclusions}
Zipper codes provide a convenient framework for describing
a wide variety of spatially-coupled product-like codes.  Such
codes are of interest in high-throughput communication systems
because they can be decoded iteratively using
low-complexity power-efficient constituent decoders while
achieving a gap to the Shannon limit on the binary
symmetric channel of 0.5~dB or less at high code rates.
We have introduced tiled diagonal and delayed diagonal interleaver
maps that couple the constituent codes in regular (hardware-friendly)
patterns.  These interleaver maps provide flexibility
in trading off decoding window size and code performance.
A combinatorial analysis of stall patterns that arise
with such interleaver maps shows that increasing
tile size or delay can indeed have a beneficial effect on
reducing the error floor of the code.

Further research on zipper codes is needed to address the design
of interleaver maps that give rise to codes with large minimum
Hamming distance or large dominant stall-pattern size.  Tradeoffs between
code performance and decoding latency should be better
characterized.  Can soft-decision (or soft-aided)
decoding methods be introduced without excessive increase
in decoding complexity and decoder power consumption?
No doubt many further interesting questions can be formulated.


\begin{IEEEbiographynophoto}{Alvin Y. Sukmadji} (S'20)
received the B.A.Sc.\ degree (with high honours) in
electrical engineering with a minor in mathematics in 2017, and the
M.A.Sc degree in electrical
and computer engineering in 2020, both from the University of Toronto, Toronto,
ON, Canada. He is currently a Ph.D. student in the Department of Electrical and
Computer Engineering at the University of Toronto. His research interests
include coding theory and fiber-optic communications. He received the NSERC
CGS-M scholarship in 2017, the Ontario Graduate Scholarship in 2018 and 2019,
and the NSERC PGS-D scholarship in 2020.
\end{IEEEbiographynophoto}

\begin{IEEEbiographynophoto}{Umberto Mart\'{i}nez-Pe\~{n}as} (S'15--M'18)
received the B.Sc. and M.Sc. degrees in Mathematics from the University of
Valladolid, Spain, in 2013 and 2014, respectively, and the Ph.D. degree in
Mathematics from Aalborg University, Denmark, in 2017. In 2018 and 2019, he was
a Post-Doctoral Fellow with the Department of Electrical and Computer
Engineering, University of Toronto, Canada. In 2020 and 2021, he was a Ma\^{i}tre
Assistant with the Institute of Computer Science and Mathematics, University of
Neuch\^{a}tel, Switzerland. His research interests include algebra, algebraic
coding, distributed storage, network coding, and information-theoretic security
and privacy. He was awarded an Elite Research Travel Grant (EliteForsk
Rejsestipendium) from the Danish Ministry of Education and Science in 2016, the
Ph.D. Dissertation Prize by the Danish Academy of Natural Sciences (Danmarks
Naturvidenskabelige Akademi, DNA) in 2018, and a Vicent Caselles Prize for
Research in Mathematics by the Royal Spanish Mathematical Society (Real
Sociedad Matem\'{a}tica Espa\~{n}ola, RSME) in 2019.
\end{IEEEbiographynophoto}

\begin{IEEEbiographynophoto}{Frank R. Kschischang} (S'83--M'91--SM'00--F'06)
received the B.A.Sc. degree (with honours) from the University of British
Columbia, Vancouver, BC, Canada, in 1985 and the M.A.Sc. and Ph.D. degrees from
the University of Toronto, Toronto, ON, Canada, in 1988 and 1991, respectively,
all in electrical engineering. Since 1991 he has been a faculty member in
Electrical and Computer Engineering at the University of Toronto.

His research interests are focused primarily on the area of channel coding
techniques, applied to wireline, wireless and optical communication systems and
networks. He has received several awards for teaching and research, including
the 2010 Communications Society and Information Theory Society Joint Paper
Award, and the 2018 IEEE Information Theory Society Paper Award.  He is a
Fellow of IEEE, of the Engineering Institute of Canada, of the Canadian Academy
of Engineering, and of the Royal Society of Canada.

During 1997--2000, he served as an Associate Editor for Coding Theory for the
\textsc{IEEE Transactions on Information Theory}, and from 2014--16, he served
as this journal's Editor-in-Chief.  He served as general co-chair for the 2008
IEEE International Symposium on Information Theory, and he served as the 2010
President of the IEEE Information Theory Society.  He received the Society's
Aaron D. Wyner Distinguished Service Award in 2016.
\end{IEEEbiographynophoto}

\end{document}